# *GliomaINK*: A Tumor and Anatomy Conditioned Flow Matching Framework for 3D MRI Sequence Synthesis in Glioma Patients

Moinak Bhattacharya PhD[1], Annie Singh MBBS[2], Mohammad Chaudhry MD[1], Prateek Prasanna PhD[3], Gagandeep Singh MD[1]

[1]Department of Radiology, Columbia University Irving Medical Center, NY, US
[2]Atal Bihari Vajpayee Institute of Medical Sciences, New Delhi, India
[3]Department of Biomedical Informatics, Stony Brook University, NY, US

**Abstract**
**Background:** Incomplete multiparametric MR examinations limit glioma characterization and downstream AI tools. Generative models can synthesize missing MRI sequences, but anatomical structure and clinically relevant tumor features are challenging to preserve. We propose *GliomaINK*, an anatomy-guided conditional flow matching framework for cross-sequence 3D brain MRI synthesis.

**Methods:** A 3D SwinUNETR predicted the conditional velocity field while an auxiliary topology-preserving loss encouraged structural consistency. The model was trained on 1,083 patients from the BraTS-AG dataset and independently tested on an in-house clinical cohort. Downstream tasks include image quality, tumor segmentation, IDH1 mutation-status prediction, and overall survival prediction. Two radiologists independently evaluated the anatomical fidelity, pathological representation, and overall image quality of the synthesized images.

**Results:** *GliomaINK* exhibited the highest PSNR across all six MRI translation directions in both evaluation cohorts. In the BraTS-AG test set, the PSNR reached 25.64 ± 0.55 dB for the T2W-to-T1C synthesis, compared with 13.78 ± 0.32 dB for Pix2Pix and 18.90 ± 0.53 dB for SPADE. For downstream tumor segmentation, the method showed strong performance across tumor subregions with Dice scores reaching 0.928 ± 0.014 for edema, and 0.885 ± 0.040 for enhancing tumor. For the IDH1 mutation-status prediction, *GliomaINK* achieved the highest AUROC score in four of the six sequence configurations, reaching 0.827 ± 0.088. In the radiologist evaluations, *GliomaINK* achieved significantly higher pathological representation scores than the baseline methods.

**Conclusion:** Anatomy-guided conditional flow matching enables high-fidelity cross-sequence glioma MRI synthesis demonstrating potential for preservation of clinically relevant information in synthesized MRI and improving downstream tasks.

## Introduction

Magnetic resonance imaging (MRI) is an essential component of neuro-oncology, providing structural, physiological, and pathological information that is critical for diagnosis, surgical planning, radiotherapy, and treatment monitoring[1,2]. Multi-parametric MRI protocols, including T1-weighted (T1), contrast-enhanced T1-weighted (T1C), T2-weighted (T2), and fluid-attenuated inversion recovery (FLAIR) sequences, provide complementary information about tumor-related processes, including blood-brain barrier disruption, edema, tumor infiltration, necrosis, and abnormalities in the surrounding tissues[3]. However, complete multiparametric MRI examinations are not always available in routine clinical practice. Individual sequences may be degraded or missing because of patient motion, time constraints, emergency imaging requirements, metal-related artifacts, variations in acquisition protocols, or contrast allergies. Such incomplete examinations can hinder the integration of complementary imaging information needed for tumor characterization and longitudinal assessment, particularly in patients with complex tumors or evolving post-treatment changes[4,5]. This limitation also affects the growing use of computational and artificial intelligence (AI)-based tools in neuro-oncology, as many methods for tumor segmentation, molecular characterization, and longitudinal disease monitoring are developed using complete and standardized multiparametric imaging inputs. Consequently, missing sequences can limit the applicability of these tools to heterogeneous real-world clinical data and contribute to a gap between performance in curated research datasets and deployment in clinical practice.

Recent advances in generative AI have emerged as a promising approach for harmonizing multimodal MRI data and reconstructing missing imaging sequences. Generative adversarial networks (GANs) have previously been used to synthesize high-quality medical images; however, these approaches are often associated with unstable training, the generation of anatomically unrealistic structures, and limited diversity in the synthesized images[6,7]. More recently, diffusion probabilistic models have demonstrated improved synthesis accuracy by modeling complex image distributions through an iterative denoising process[8–13]. Diffusion models have been successfully applied to a range of medical imaging tasks, including cross-sequence image translation, super-resolution, image restoration, and MRI synthesis. Despite their potential, diffusion-based methods are computationally intensive because they typically require multiple denoising steps during image generation. Furthermore, both GAN- and diffusion-based approaches generally rely on the assumption that the intensity information contained in the available MRI sequences is sufficient to generate anatomically accurate images, which may not always be the case when reconstructing missing sequences.

Conditional flow matching (CFM) has emerged as a promising alternative to conventional diffusion-based generative models[14–16]. Unlike diffusion models, which generate images by reversing a stochastic noising process, flow matching learns continuous probability paths between data distributions through differential equations. This formulation enables faster sampling while maintaining high image quality and stable optimization, and recent studies have demonstrated that flow-based approaches can achieve improved performance to established generative methods. These advantages make CFM an attractive framework for efficient MRI synthesis and reconstruction of missing sequences. Nevertheless, an important challenge remains in generating anatomically faithful and structurally consistent brain MRI. While CFM offers advantages in sampling efficiency and generative modeling, emerging flow-based approaches have primarily focused on conditioning generation on image information without explicitly incorporating anatomical representations or topological constraints. The anatomical structure of the brain is considerably more complex than can be represented by imaging information alone. Key anatomical structures, including white matter pathways, cortical and deep gray matter, ventricular structures, cerebrospinal fluid spaces, and tumor regions, exhibit spatial relationships that remain relatively consistent across individuals. At the same time, brain tumors introduce complex anatomical deformations resulting from mass effect, infiltration, edema, necrosis, and postoperative changes, leading to nonlinear interactions between healthy anatomy and disease-related abnormalities. Consequently, synthesized images may contain subtle anatomical inaccuracies or unrealistic tissue structures, highlighting the importance of incorporating explicit anatomical guidance to ensure reliable and clinically meaningful image generation. We hypothesize that explicit anatomical conditioning on tissue structures and tumor morphology can disentangle structural information from sequence-specific intensity distributions, thereby improving the anatomical fidelity and biological relevance of synthesized MRI.

To summarize, we present *GliomaINK*, an anatomy-guided framework for generating multiparametric MRI sequences in neuro-oncology. *GliomaINK* integrates three complementary sources of information: brain anatomic representations, tumor segmentation masks, and available MRI sequences, together with a topology-preserving objective to promote anatomically faithful synthesis. We evaluate *GliomaINK* on the BraTS dataset and an independent clinical cohort with heterogeneous imaging protocols, comparing its performance with established GAN- and diffusion-based methods. We further assess the clinical utility of the generated images through downstream tasks, including tumor segmentation, molecular biomarker prediction, and overall survival estimation, as well as qualitative evaluation by three radiologists assessing anatomical fidelity, pathological representation, and overall image quality.

## Material and methods

### Study Design and Overview

Our team has established a generative AI framework employed in analyzing brain MRI in adults with gliomas. The model was trained on 1,083 patients from the BraTS Adult Glioma (BraTS-AG) dataset[17]. We conducted external validation with the help of an independent test set made up of 112 patients from the BraTS-AG dataset and in-house cohort of 272 patients[18]. We use the three sequences that are diagnostically important T1-weighted contrast-enhanced (T1C), T2-weighted (T2W), and T2-FLAIR (T2F). The anatomical structures that were used as conditioning inputs include white matter pathways (WMT), cortical gray matter (CGM), lateral ventricles (LV), and tumor segmentation masks.

### Study Population

The Institutional Review Board (IRB) granted approval for conducting this retrospective study (IRB Approval No. IRB-AAAJ9652). The patients diagnosed with glioma between 2010 and 2017 were identified as eligible candidates to be included in the study where MRI scans had all three required sequences (T1C, T2W, and T2F) available. IDH mutation status was collected from available clinical records and pathology reports, where documented, and used as the reference label for downstream prediction analyses.

### Anatomy-guided Conditional Flow Matching Model and Training

We designed an anatomy-guided conditional flow matching (CFM) model to synthesize the target MRI sequence from an available input sequence along with the anatomical and tumor segmentation masks. The conditioning representation was formed by concatenating the input image, a three-channel anatomical representation (WMT, CGM and LV), and a three-channel tumor representation (ET, NC and PE), resulting in six conditioning channels. The target image was paired with Gaussian noise, and an intermediate state was sampled using a linear interpolation, $x_t=(1-t)*x_0+t*z$, where t was sampled uniformly from 0 to 1. The model predicted the conditional velocity field using a 3D SwinUNETR[19] with one input image channel plus five conditioning channels, while the sampled time variable was provided to the model. The flow-matching objective minimized the mean-squared error between the predicted velocity and the target velocity $x_0$-z. To encourage preservation of structural topology, an auxiliary topology loss based on iterative soft morphological erosion was added with a weighting factor of 0.01. The total training objective was therefore $L=L_{flow}+0.01L_{topo}$. The model was optimized using Adam with a learning rate of $2\times10^{-4}$, with training performed for up to 500 epochs and a batch size of two. During inference, a Gaussian-noise volume was progressively transformed using 20 Euler integration steps, with the learned velocity field evaluated at each time point. The input image size was 160 × 192 × 64 with 1 mm isotropic spacing (RAS orientation) and each model was trained on a 46 GB Quadro RTX 8000 GPU.

**Tumor Segmentation**

Tumor segmentation was performed using nnU-Net[20], an automated deep learning framework for biomedical image segmentation, with preprocessing and model configuration parameters automatically selected by the nnU-Net pipeline. The model was trained to identify the major glioma subregions, including peritumoral edema (PE), necrotic tumor core (NC), and enhancing tumor (ET). We defined six different combinations namely, real T1C with synthetic T2F and T2W, real T2F with synthetic T1C and T2W, real T2W with synthetic T1C and T2F, real T1C and T2F with synthetic T2W, real T1C and T2W with synthetic T2F, and real T2F and T2W with synthetic T1C. The nnU-Net was trained separately on these different combinations, and the glioma segmentation masks were generated.

**IDH1 and Overall Survival Prediction**

The clinical relevance of the synthesized MRI sequences was evaluated by its performance on downstream tasks like IDH1 mutation prediction and OS analysis. We extracted the radiomics features using PyRadiomics[21]. We use the six different combinations, discussed in the previous subsection, and 1409 radiomic features encompassing shape, first-order intensity, and texture characteristics were extracted from the tumor regions.

For IDH1 mutation-status prediction, we used the extracted radiomic features in a binary classification pipeline. We first standardize the features and then the top-k were selected using a univariate feature selection method with an ANOVA F-test retaining the 20 highest-ranked features. We trained a Random Forest classifier using the top 20 selected features and evaluated performance using the area under the receiver operating characteristic curve (AUROC). Similarly, for OS analysis, the features were initially standardized and reduced to 10 principal components using principal component analysis (PCA). We then fitted a penalized Cox proportional hazards model with a penalization parameter of 0.5 and evaluated performance using concordance index (C-index). For both downstream tasks, we evaluated the entire preprocessing and modeling pipeline independently using five-fold stratified cross-validation. Four folds were used for model development and one-fold was held out for evaluation for each five-fold split, with each fold serving as the held-out fold once. This yielded five-fold AUROC values for IDH1 prediction and C-index values for OS prediction for each experimental condition. We summarize the results as mean ± standard deviation across the five folds.

**Radiologist Evaluation**

Two radiologists independently evaluated the generated MRI sequences to assess their clinical realism and diagnostic utility. Both radiologists evaluated the images using three predefined scoring criteria: anatomical fidelity (0-4), pathological representation (0-4), and overall image quality (0-2). The detailed scoring mechanism is shown in Supplementary Table 1. Anatomical fidelity assessed whether the generated images accurately preserved normal brain structures, while pathological representation evaluated the visualization and characteristics of the tumor and its relevant subregions. Overall image quality was assessed based on factors such as image clarity, artifacts, contrast, and visual realism. The scores from both the radiologists were aggregated to compare the quality and clinical plausibility of generated images across different MRI sequences and missing-sequence reconstruction settings.

### Statistical Analysis

All statistical analyses were performed to evaluate both image synthesis quality and downstream clinical task performance. For image synthesis, we computed standard metrics including Structural Similarity Index (SSIM) and Peak Signal-to-Noise Ratio (PSNR) to compare synthesized MRI sequences against ground truth. Segmentation quality was evaluated against the available reference masks using standard metrics such as the Dice similarity coefficient (DSC) and 95% percentile Hausdorff distance (HD95). For IDH1 prediction, classification performance was assessed using AUROC and OS analysis was evaluated using C-index. All the experiments we compared against two reference methods Pix2Pix[22] and SPADE[23]. Statistical significance between groups was tested using paired t-tests or Wilcoxon signed-rank tests, depending on normality assumptions. All analyses were conducted using Python (Scikit-learn, Lifelines) with a significance threshold set at $p < 0.05$.

## Results

### Patient Characteristics

The in-house cohort consisted of a total of 292 patients (Table 1). Of those, 166 patients (56.8%) were male, while 113 patients (38.7%) were female. The sex variable was unreported for 13 patients (4.5%). In terms of IDH mutation status, 255 patients (87.3%) were reported as wildtype, 25 patients (8.6%) were mutant, while the status was unknown for 12 patients (4.1%). As for the survival status, 152 patients (52.1%) were dead, 29 patients (9.9%) were alive or censored, while the survival status was unknown for 111 patients (38%).

### Anatomy-aware conditioning improves high-fidelity MRI reconstruction

The proposed *GliomaINK* generated the best quantitative MRI sequence across the six different combinations (Table 2). We use both the BraTS-AG dataset and in-house cohort for evaluation. When compared to Pix2Pix and SPADE on the BraTS-AG dataset, the proposed method achieved the highest PSNR in all six translation directions. The largest improvement over Pix2Pix was observed for the T2W to T1C translation, where PSNR increased from 13.78 ± 0.32 dB with Pix2Pix to 25.64 ± 0.55 dB with the proposed method. Similarly, for the T2F to T1C translation, PSNR increased from 13.49 ± 0.28 dB to 24.95 ± 0.51 dB. In comparison to SPADE, the proposed method achieved higher PSNR than SPADE in all six translation tasks, with improvements ranging from 0.67 dB for T1C to T2W to 6.74 dB for T2W to T1C translation. Across all six translation directions, the mean PSNR was 16.94 dB for Pix2Pix, 20.33 dB for SPADE, and 23.42 dB for the proposed method. The proposed method also yielded favorable structural similarity. It achieved the highest SSIM in four out of the six translation tasks: T1C to T2F (0.755 ± 0.008), T2F to T1C (0.782 ± 0.008), T2W to T1C (0.805 ± 0.008), and T2W to T2F translation (0.775 ± 0.010). Pix2Pix achieved the highest SSIM for T1C to T2W (0.794 ± 0.013) and T2F to T2W translation (0.786 ± 0.013), although the difference between Pix2Pix and the proposed method for T2F to T2W was small. The mean SSIM across the six translation tasks was 0.72 for Pix2Pix, 0.66 for SPADE, and 0.78 for the proposed method.

Qualitative assessment further demonstrated that the anatomy-aware approach produced synthetic images that were closer to the target MRI anatomy and visible pathological features (Figure 2). Compared to Pix2Pix and SPADE, the proposed method more consistently preserved the shape and spatial characteristics of relevant structures across T1C, T2W, and T2F images. These quantitative and qualitative findings demonstrate that anatomical conditioning improves the fidelity of cross-modal MRI synthesis.

The performance of the synthesis methods was further evaluated on the CUIMC external validation cohort, using the same evaluation methodology. Similar to the performance reported for the BraTS-AG dataset, the proposed method achieved the highest PSNR across all six translation directions, with values ranging from 21.55 ± 0.31 dB for T1C to T2F to 23.21 ± 0.36 dB for T2W to T1C translation. The largest PSNR advantage over Pix2Pix was observed for the T2W to T1C translation (23.21 ± 0.36 vs. 13.34 ± 0.32 dB). For SSIM, the proposed method achieved the highest value in four of the six translation tasks: T2F to T1C (0.747 ± 0.006), T2W to T1C (0.759 ± 0.006), and T2W to T2F translation (0.718 ± 0.008), with Pix2Pix achieving the highest SSIM for T1C to T2F, T1C to T2W, and T2F to T2W. Overall, the external validation results demonstrate that the improved PSNR performance of the proposed approach was retained in an independent cohort, while SSIM performance remained competitive across translation directions.

**Anatomy-aware conditioning improves tumor segmentation across MRI contrasts**

In this subsection, we discuss the application of the synthesized MRI sequences for downstream tumor segmentation on the BraTS-AG dataset (Table 3). The proposed approach consistently achieved high segmentation performance across evaluated real/synthetic-channel configurations and tumor compartments.

For edema, the proposed approach achieved particularly high Dice scores when T2F or T2W were the real anchor channel. With T2F as real channel and the remaining two modalities synthesized, the proposed approach achieved 0.885 ± 0.040, versus 0.522 ± 0.079 and 0.516 ± 0.069 for Pix2Pix and SPADE, respectively. For T2W as real channel, the corresponding scores were 0.881 ± 0.032, 0.547 ± 0.079, and 0.557 ± 0.067, respectively. When T2F and T2W were real and T1C synthesized, the proposed approach achieved 0.833 ± 0.046 versus 0.581 ± 0.071 and 0.579 ± 0.074 for Pix2Pix and SPADE. For enhancing tumor, the proposed approach achieved the highest Dice score in most configurations, with the largest performance differences when a single T1C, T2F, or T2W channel was used as real anchor. For example, with T1c as real channel, the proposed approach achieved 0.893 ± 0.012 versus 0.729 ± 0.049 and 0.720 ± 0.046 for Pix2Pix and SPADE, respectively. T1C and T2W as real channels had corresponding enhancing tumor Dice scores of 0.896 ± 0.016 for the proposed approach versus 0.831 ± 0.042 and 0.830 ± 0.041 for Pix2Pix and SPADE. For necrotic core, the proposed approach obtained the highest Dice score across all six configurations. Scores varied between 0.860 ± 0.011 for the T1C+T2F condition and 0.928 ± 0.014 when T2F was the real anchor channel. In the latter, Pix2Pix and SPADE achieved Dice scores of 0.554 ± 0.038 and 0.554 ± 0.031, respectively. Similarly, for T2W as real anchor, the proposed approach achieved 0.927 ± 0.016 versus 0.552 ± 0.051 and 0.559 ± 0.044 for Pix2Pix and SPADE.

The all-real reference configuration achieved Dice scores of 0.781 ± 0.055, 0.865 ± 0.029, and 0.854 ± 0.011 for edema, enhancing tumor, and necrotic core, respectively. Across several configurations, the proposed approach achieved segmentation performance comparable to or better than the all-real reference, particularly when T2F or T2W were the real anchor. This suggests that the anatomical information preserved during synthesis can be utilized to achieve downstream tumor segmentation, even with one or more synthesized MRI contrasts.

Qualitative evaluation showed similar findings (Figure 3). The segmentations generated with the proposed approach were closer to the visible tumor boundaries and spatial distributions of the enhancing and peritumoral compartments. Interestingly, the predicted tumor regions were still spatially consistent with the underlying lesion, despite differences in MRI contrast, suggesting that the quantitative evidence of improved preservation of clinically relevant anatomical information translates to consistent anatomical segmentation.

**Anatomically Faithful Synthesis Preserves Clinically Relevant Information**

To determine whether the synthesized images preserved information beyond visual and anatomical similarity, we evaluated their utility for downstream molecular characterization using IDH1 prediction and OS analysis on the in-house dataset (Table 4). Across the different configurations, the proposed method generally achieved competitive or superior downstream performance compared to Pix2Pix and SPADE.

For IDH1 prediction, the proposed method achieved the highest AUC in four of the six configurations. The highest AUC was achieved with T2W as the real anchor (AUC= 0.827 ± 0.088 ($p = 0.002$)). The proposed method also achieved the highest AUC for T2F (AUC= 0.727 ± 0.101 ($p = 0.011$)), T1C+T2F (AUC= 0.737 ± 0.143 ($p = 0.030$)), and T2F+T2W (AUC= 0.777 ± 0.085 ($p = 0.003$)). For the T1c and T1C+T2W configurations, Pix2Pix achieved higher AUC values compared to the proposed method (AUC= 0.785 ± 0.118 and 0.808 ± 0.115, respectively).

For overall-survival prediction, the proposed method achieved the highest concordance index in the T2F, T1C+T2F, and T2F+T2W configurations (C-index= 0.624 ± 0.027, 0.610 ± 0.072, and 0.591 ± 0.065, respectively). The proposed method also achieved the highest C-index in the T1C+T2W configuration compared with the other methods. In the T2W configuration, SPADE achieved the highest C-index (C-index= 0.576 ± 0.082), whereas the proposed method achieved 0.557 ± 0.052. Thus, the survival results were more heterogeneous compared to the IDH1 results.

Taken together, the downstream analyses demonstrate that the proposed anatomy-aware synthesis strategy can preserve information relevant to molecular and survival prediction. The strongest and most consistent benefit was observed for the IDH1 prediction, while the survival prediction was configuration-dependent across the synthesis methods. These findings complement the image-quality and segmentation analyses by showing that anatomical fidelity can be associated with preservation of information useful for downstream clinical prediction.

### Radiologists' evaluation

Figure 4 summarizes the combined radiologist evaluation results, showing that *GliomaINK* achieved the highest pathological fidelity scores in both evaluations (3.62 ± 0.56 and 3.70 ± 0.46), compared with SPADE (2.46 ± 0.89 and 3.52 ± 0.56) and Pix2Pix (2.42 ± 0.85 and 3.45 ± 0.67). A major finding of the reader evaluation was that *GliomaINK*-generated T1-weighted post-contrast (T1C) images received significantly higher ratings for pathological depiction and overall scores than the comparator models. This finding is particularly noteworthy because *GliomaINK* synthesized these images from non-contrast T2-weighted input alone, without requiring a contrast-enhanced acquisition. Although previous studies have demonstrated the feasibility of synthesizing contrast-enhanced MRI from non-contrast sequences, reliable depiction of lesion enhancement remains challenging, with reported failures including missed or underestimated enhancing regions[24]. *GliomaINK*'s superior reader-rated pathological depiction therefore supports further investigation of T2W-to-T1C synthesis for selected clinical tasks in which reducing reliance on gadolinium-based contrast agents would be desirable. *GliomaINK* also maintained high anatomical fidelity, achieving scores of 3.94 ± 0.24 and 3.95 ± 0.22, while image quality remained comparable across the three models. The individual radiologist scores are provided in Supplementary Tables 2 and 3 provide the radiologist-specific scores. Figure 5 illustrates inter-radiologist variability in scoring the same patient cases, with one radiologist assigning higher scores and another lower scores, highlighting the subjective nature of visual assessment of disease-related features and anatomical preservation. The failure cases in Figure 6 highlight discrepancies between conventional quantitative image-quality metrics and radiologist assessments, underscoring the limitations of these metrics in capturing clinically relevant pathological and anatomical details. Taken together, these findings highlight the central strength of *GliomaINK* in improving pathological fidelity without sacrificing anatomical consistency, while also revealing the remaining challenges in generating clinically convincing images.

## Discussions

In this work, we propose *GliomaINK*, an anatomy-aware framework for cross-sequence brain MRI synthesis trained on the BraTS-AG training dataset with the aim of addressing the problem of limited multi-parametric MRI acquisitions. The main finding was that the use of anatomical conditioning resulted in the enhanced image fidelity across MRI translation tasks, and more importantly, preserved the information critical for downstream tumor segmentation and clinical prediction. The benefit of anatomy guided synthesis was most prominent for PSNR and tumor segmentation, with comparable gains for SSIM on the BraTS-AG test set. The method was also evaluated on an in-house cohort for independent validation for the image quality assessment, IDHH1 prediction and OS analysis the observed gains were consistent with the BraTS results.

The image synthesis results broadly support the value of anatomical information for MRI reconstruction beyond standard image appearance conditioning. *GliomaINK* achieved the highest PSNR across all six translation directions on both the BraTS test set and the in-house cohort. The largest gains were observed in the translations targeting T1C, specifically T2W to T1C and T2F to T1C translation. This is an important result since T1C is a challenging contrast due to blood-brain barrier opening and thus its appearance is only partially informative about tumor location and edema. By explicitly leveraging anatomical information about the tumor and tissue structures, *GliomaINK* can better constrain the generation process and avoid unrealistic deformations during T1C synthesis in the absence of that sequence. Notably, the external validation on the in-house cohort showed similarly strong performance gains, which was not trivial since that cohort was acquired on a different scanner with a more varied acquisition protocol. The SSIM results provide an additional perspective since they highlight the trade-offs between perceptual similarity of the synthesized images to the real MRI scans. While *GliomaINK* achieved the highest SSIM in four out of six combinations on the BraTS test set, Pix2Pix was better in T1C to T2W and T2F to T2W translation. The same was observed in the in-house cohort with Pix2Pix having an edge in terms of perceptual similarity. Notably, while Pix2Pix achieved higher SSIM in certain directions, it did not result in higher Dice scores in any of the segmentations.

The downstream segmentation results highlight the ability of *GliomaINK* to predict tumor masks from MRI scans that were partially synthesized from other modalities. *GliomaINK* reported the highest Dice scores across all six configurations for the necrotic core while also significantly outperforming the other methods for edema and enhancing. In several cases, the performance of segmentations from the synthesized MRI channels were comparable to the all-real reference. This is an important finding since it suggests that the quality of image synthesis should not be assessed in isolation from the ability to perform downstream tasks. In particular, there are cases when a tumor segmentation from a synthesized MRI scan can be better than a segmentation of the same real scan, potentially due to the removal of artifacts affecting tumor delineation.

The IDH1 prediction and OS analysis tasks demonstrated that the preserved tumor information in the synthesized MRI scans translated into equivalent performance on clinically relevant prediction tasks. *GliomaINK* achieved the best AUROC performance for IDH1 prediction in four out of six sequence configurations. While Pix2Pix performed better than *GliomaINK* in the T1C and T1C+T2W configurations for IDH prediction, these were the only configurations where the tumor information was preserved in the synthesized scans. The comparable performance of *GliomaINK* in the other four configurations highlights its ability to preserve radiomic features critical for IDH1 prediction outperforming the best baseline significantly ($p<.05$).

The radiologist evaluation further supported these quantitative findings, with *GliomaINK* achieving the highest combined pathological fidelity scores across both evaluations while maintaining anatomical fidelity comparable to SPADE and Pix2Pix. Notably, this advantage was most pronounced in the first evaluation, where *GliomaINK* achieved a pathological fidelity score of $3.62 \pm 0.56$, compared with $2.46 \pm 0.89$ for SPADE and $2.42 \pm 0.85$ for Pix2Pix. These findings suggest that the benefit of anatomical conditioning extends beyond pixel-level similarity and may improve the preservation of clinically relevant pathological features, although image quality did not show a consistent advantage across evaluations.

The main limitation of this work is that the model is only trained for glioma patients. Downstream analysis was also limited to tumor segmentation, IDH status, and OS analysis. Other clinically relevant biomarkers, longitudinal modeling, and additional downstream clinical tasks were not investigated and should be explored in future studies. Finally, while the external validation on the in-house cohort provides proof-of-concept for the benefit of incorporating anatomical information into image synthesis, additional multi-center studies are needed to determine the generalizability of the proposed approach across different scanners, imaging protocols, and patient populations.

In summary, the results of this work demonstrate that anatomical information can be leveraged to enhance the quality of cross-sequence brain MRI synthesis. *GliomaINK* consistently achieved better image fidelity and perceptual quality compared to the baseline methods. The preserved tumor information enabled similar tumor segmentation performance from synthesized images to the tumor segmentation from real images as well as comparable performance on downstream radiomics-based prediction tasks. The external validation on the in-house cohort demonstrated that the benefits of anatomical conditioning were not specific to the BraTS dataset but were present in an independent clinical cohort as well. The variable performance across different modalities and downstream tasks highlights the importance of evaluating MRI synthesis in the context of specific applications.

Table 1: Patient and clinical characteristics of the in-house glioma cohort.

| Characteristic | Category | N |
|---|---|---|
| **Gender** | Male | 166 |
| | Female | 113 |
| | NA | 13 |
| **IDH mutation status** | Wildtype | 255 |
| | Mutant | 25 |
| | NA | 12 |
| **Overall survival status** | Deceased | 152 |
| | Alive/Censored | 29 |
| | NA | 111 |

Table 2: Quantitative evaluation of cross- sequence MRI synthesis on the BraTS-AG test set and independent in-house cohort. Best and second-best results (mean±std (95% CI)) are in bold and italics, respectively.

| **Cohort** | **Translation** | **PSNR (dB)** | | | **SSIM** | | |
|---|---|---|---|---|---|---|---|
| | | **Pix2Pix** | **SPADE** | **Ours** | **Pix2Pix** | **SPADE** | **Ours** |
| **BraTS-AG (n=116)** | T1C→T2F | 16.60 ± 0.53 | *17.29 ± 0.46* | **21.99 ± 0.49** | *0.713 ± 0.013* | 0.531 ± 0.011 | **0.755 ± 0.008** |
| | T1C→T2W | 20.60 ± 0.43 | *22.18 ± 0.35* | **22.92 ± 0.44** | **0.794 ± 0.013** | 0.735 ± 0.009 | *0.791 ± 0.011* |
| | T2F→T1C | 13.49 ± 0.28 | *22.70 ± 0.56* | **24.95 ± 0.51** | 0.619 ± 0.011 | *0.690 ± 0.009* | **0.782 ± 0.008** |
| | T2F→T2W | 19.71 ± 0.44 | *21.66 ± 0.34* | **22.85 ± 0.36** | **0.786 ± 0.013** | 0.695 ± 0.008 | *0.782 ± 0.010* |
| | T2W→T1C | 13.78 ± 0.32 | *18.90 ± 0.53* | **25.64 ± 0.55** | 0.656 ± 0.012 | *0.685 ± 0.010* | **0.805 ± 0.008** |
| | T2W→T2F | 17.45 ± 0.61 | *19.26 ± 0.60* | **22.14 ± 0.51** | *0.734 ± 0.016* | 0.654 ± 0.012 | **0.775 ± 0.010** |
| **CUMC (n=292)** | T1C→T2F | 17.91 ± 0.36 | *20.01 ± 0.26* | **21.55 ± 0.31** | **0.721 ± 0.008** | 0.600 ± 0.008 | *0.708 ± 0.007* |
| | T1C→T2W | 20.06 ± 0.30 | *21.21 ± 0.31* | **21.71 ± 0.29** | **0.750 ± 0.009** | 0.724 ± 0.009 | 0.743 ± 0.008 |
| | T2F→T1C | 12.92 ± 0.29 | *21.90 ± 0.39* | **23.11 ± 0.37** | 0.621 ± 0.008 | *0.690 ± 0.008* | **0.747 ± 0.006** |
| | T2F→T2W | 19.52 ± 0.33 | *20.82 ± 0.29* | **21.66 ± 0.29** | **0.747 ± 0.010** | 0.680 ± 0.008 | *0.730 ± 0.007* |
| | T2W→T1C | 13.34 ± 0.32 | *18.84 ± 0.38* | **23.21 ± 0.36** | 0.661 ± 0.009 | *0.677 ± 0.008* | **0.759 ± 0.006** |
| | T2W→T2F | 17.82 ± 0.39 | *20.17 ± 0.32* | **21.66 ± 0.32** | **0.732 ± 0.011** | 0.661 ± 0.009 | *0.718 ± 0.008* |

Table 3. Glioma tumor segmentation performance using real and synthesized MRI sequences. DSC (mean±std (95% CI)) score is reported with best and second-best results are in bold and italics respectively.

| Method | | ET | p | NC | p | PE | p |
|---|---|---|---|---|---|---|---|
| | Pix2Pix | *0.781* ± *0.061* | | *0.729* ± *0.049* | 0.0011 | 0.843 ± 0.014 | |
| ● ○ ○ | SPADE | **0.785 ± 0.062** | - | 0.720 ± 0.046 | | *0.851* ± *0.010* | 0.0112 |
| | **Ours** | 0.780 ± 0.061 | | **0.893 ± 0.012** | | **0.865 ± 0.010** | |
| | Pix2Pix | *0.522* ± *0.079* | | 0.801 ± 0.032 | 7.81e-05 | *0.554* ± *0.038* | |
| ○ ● ○ | SPADE | 0.516 ± 0.069 | 5.42e-05 | *0.804* ± *0.028* | | 0.554 ± 0.031 | 4.44e-05 |
| | **Ours** | **0.885 ± 0.040** | | **0.883 ± 0.020** | | **0.928 ± 0.014** | |
| | Pix2Pix | 0.547 ± 0.079 | | *0.774* ± *0.042* | 8.99e-04 | 0.552 ± 0.051 | |
| ○ ○ ● | SPADE | *0.557* ± *0.067* | 4.64e-04 | 0.769 ± 0.042 | | *0.559* ± *0.044* | 1.52e-04 |
| | **Ours** | **0.881 ± 0.032** | | **0.903 ± 0.011** | | **0.927 ± 0.016** | |
| | Pix2Pix | **0.783 ± 0.060** | | *0.860* ± *0.034* | 0.0017 | **0.862 ± 0.019** | |
| ● ● ○ | SPADE | 0.778 ± 0.062 | - | 0.859 ± 0.037 | | 0.857 ± 0.014 | - |
| | **Ours** | *0.782* ± *0.050* | | **0.889 ± 0.027** | | *0.860* ± *0.011* | |
| | Pix2Pix | **0.786 ± 0.062** | | *0.831* ± *0.042* | 0.0077 | 0.848 ± 0.011 | |
| ● ○ ● | SPADE | 0.779 ± 0.074 | - | 0.830 ± 0.041 | | *0.855* ± *0.012* | 0.0492 |
| | **Ours** | *0.782* ± *0.061* | | **0.896 ± 0.016** | | **0.867 ± 0.014** | |
| | Pix2Pix | *0.581* ± *0.071* | | *0.819* ± *0.032* | 2.50e-04 | 0.604 ± 0.042 | |
| ○ ● ● | SPADE | 0.579 ± 0.074 | 6.41e-04 | 0.817 ± 0.031 | | *0.609* ± *0.039* | 3.45e-04 |
| | **Ours** | **0.833 ± 0.046** | | **0.875 ± 0.022** | | **0.892 ± 0.025** | |
| ● ● ● | - | 0.781 ± 0.055 | | 0.865 ± 0.029 | | 0.854 ± 0.011 | |

Table 4. Performance of downstream IDH1 mutation and overall survival (OS) prediction using real and synthesized MRI sequences on the in-house dataset.

| | Method | IDH1 (AUC) | *p* | OS (C-index) | *p* |
|---|---|---|---|---|---|
| | Ours | 0.725 ± 0.144 | | 0.582 ± 0.038 | |
| ● ○ ○ | Pix2Pix | **0.785 ± 0.118** | - | **0.606 ± 0.075** | - |
| | SPADE | 0.753 ± 0.141 | | 0.592 ± 0.026 | |
| | Ours | **0.727 ± 0.101** | | **0.624 ± 0.027** | |
| ○ ● ○ | Pix2Pix | 0.566 ± 0.085 | 0.011 | 0.578 ± 0.049 | 0.001 |
| | SPADE | 0.521 ± 0.230 | | 0.617 ± 0.058 | |
| | Ours | **0.827 ± 0.088** | | 0.557 ± 0.052 | |
| ○ ○ ● | Pix2Pix | 0.633 ± 0.208 | 0.002 | 0.569 ± 0.069 | 0.095 |
| | SPADE | 0.666 ± 0.124 | | **0.576 ± 0.082** | |
| | Ours | **0.737 ± 0.143** | | **0.610 ± 0.072** | |
| ● ● ○ | Pix2Pix | 0.723 ± 0.112 | 0.030 | 0.602 ± 0.066 | 0.038 |
| | SPADE | 0.736 ± 0.153 | | 0.603 ± 0.080 | |
| | Ours | 0.773 ± 0.146 | | 0.569 ± 0.076 | |
| ● ○ ● | Pix2Pix | **0.808 ± 0.115** | - | 0.568 ± 0.080 | - |
| | SPADE | 0.762 ± 0.174 | | **0.587 ± 0.059** | |
| | Ours | **0.777 ± 0.085** | | 0.591 ± 0.065 | |
| ○ ● ● | Pix2Pix | 0.616 ± 0.112 | 0.003 | 0.574 ± 0.063 | - |
| | SPADE | 0.612 ± 0.188 | | **0.610 ± 0.051** | |

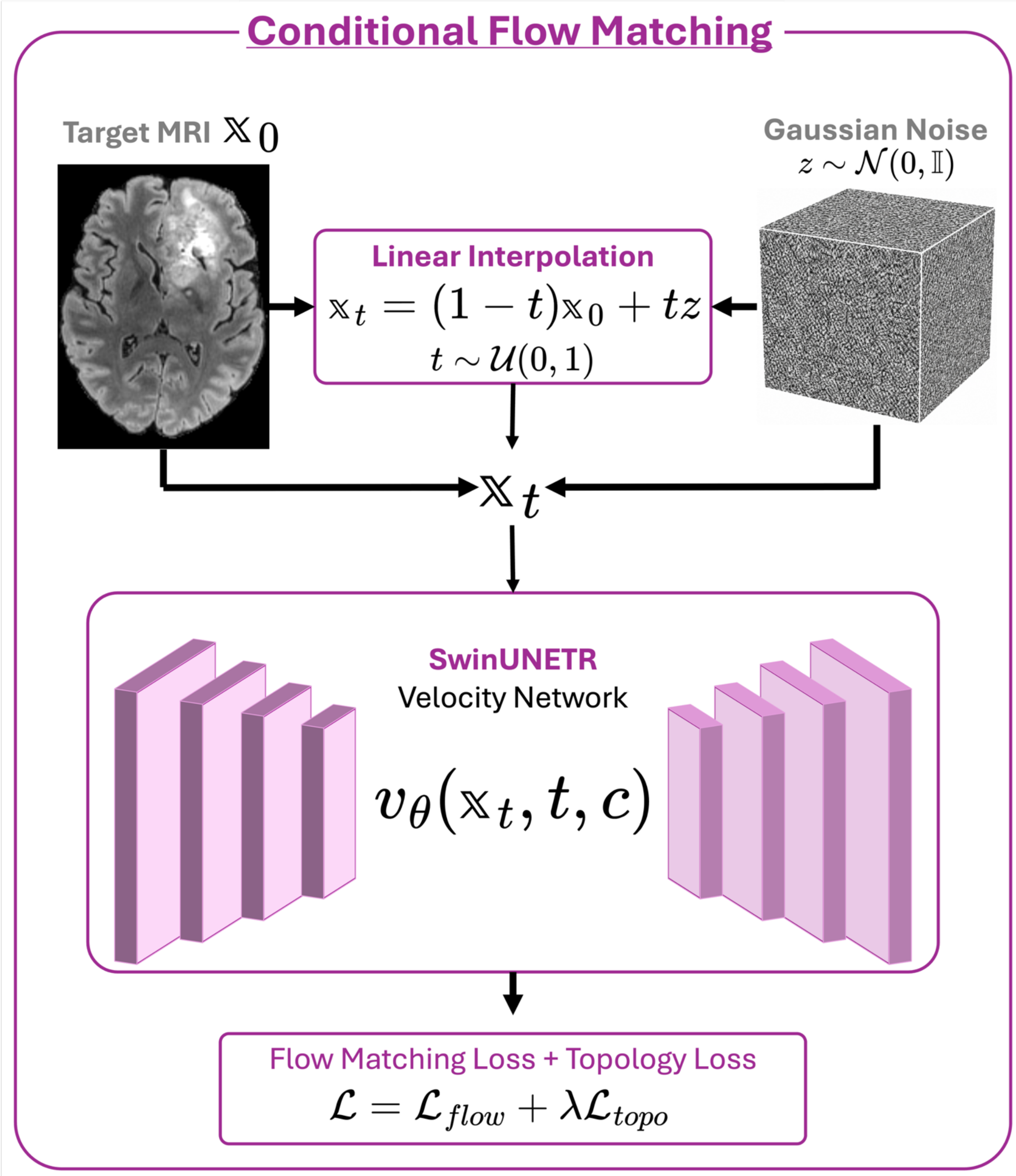

Figure 1: *Overview of the GliomaINK anatomy-guided conditional flow matching framework for 3D brain MRI synthesis.* The model takes an available MRI sequence, anatomical representations of white matter tracts, cortical gray matter, and lateral ventricles, and tumor subregion masks as conditioning inputs. A 3D SwinUNETR predicts the conditional velocity field to transform Gaussian noise into a synthesized target MRI sequence through iterative Euler integration. An auxiliary topology-preserving loss is incorporated into the training objective to encourage preservation of anatomical structure.

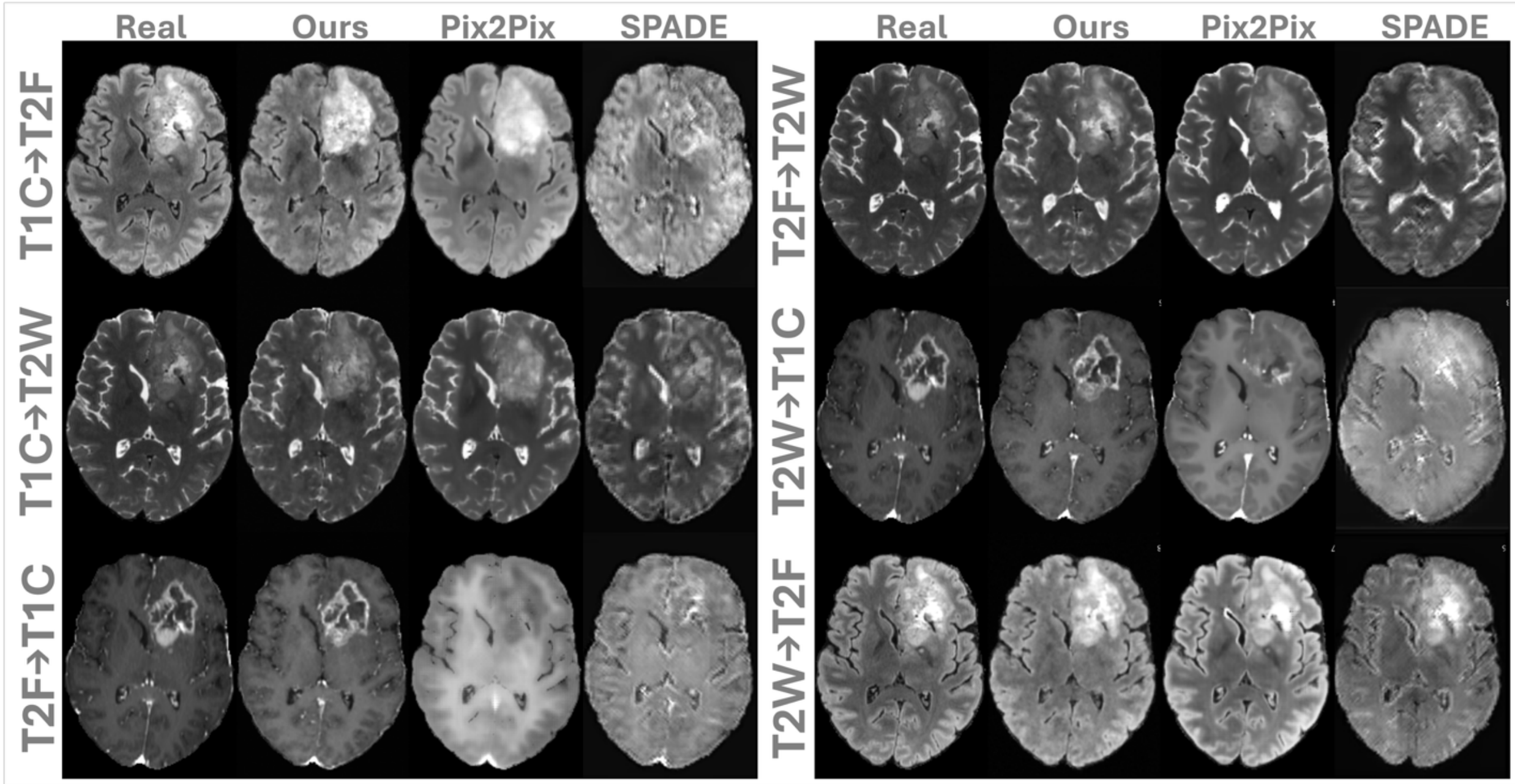

Figure 2: *Qualitative comparison of cross-sequence MRI synthesis by GliomaINK, Pix2Pix, and SPADE.* Representative examples illustrate the synthesis of T1-weighted contrast-enhanced (T1C), T2-weighted (T2W), and T2-FLAIR (T2F) images from the available MRI sequence. Synthesized images are compared with the corresponding ground-truth images to assess preservation of brain anatomy and tumor-related features. Quantitative performance is summarized using PSNR and SSIM for six translation directions in the BraTS-AG test set.

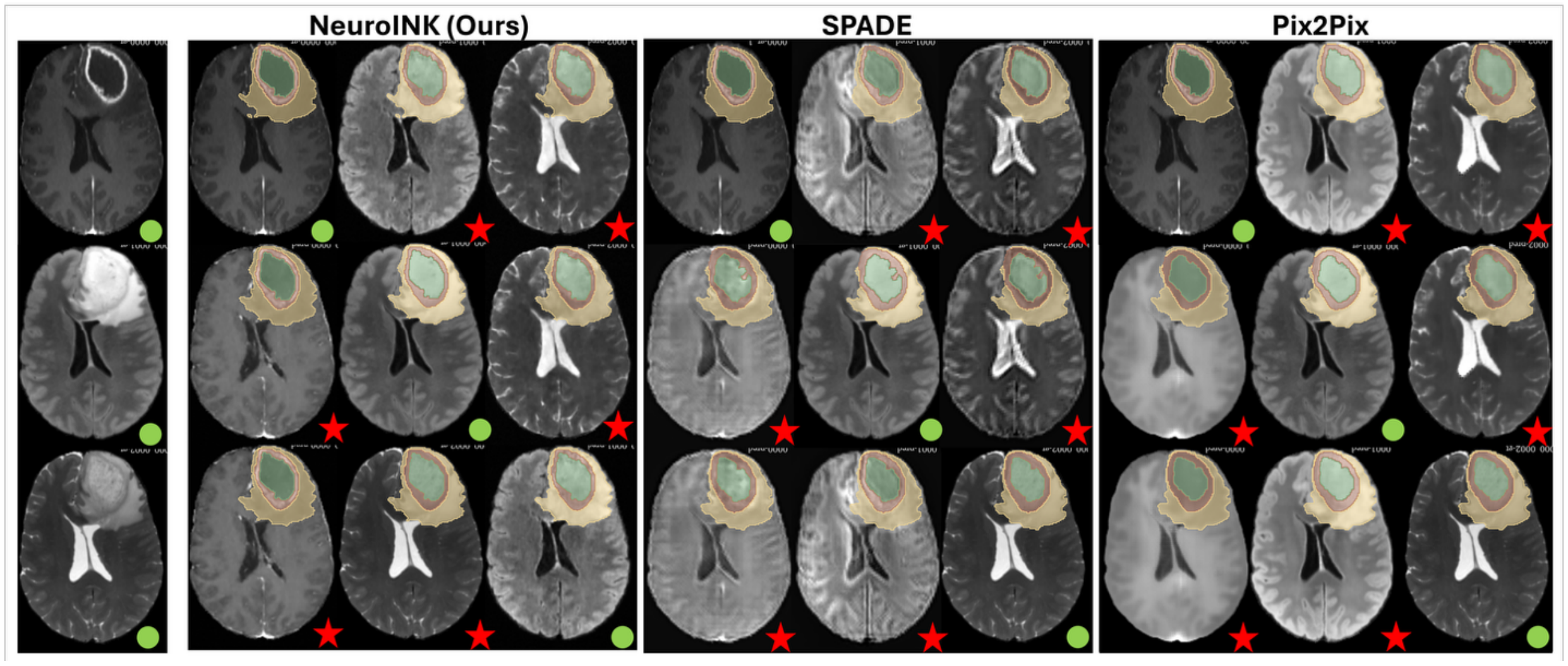

Figure 3: *Comparison of glioma tumor segmentation using real and synthesized MRI sequences.* Representative segmentation results obtained using nnU-Net with MRI sequence combinations containing real and synthesized sequences generated by *GliomaINK*, Pix2Pix, and SPADE. Tumor subregions include enhancing tumor (ET in red), necrotic core (NC in green), and peritumoral edema (PE in yellow). Predicted segmentation masks are compared with reference annotations to illustrate differences in tumor boundary delineation and spatial distribution across the evaluated configurations. Quantitative segmentation performance is reported using the Dice similarity coefficient.

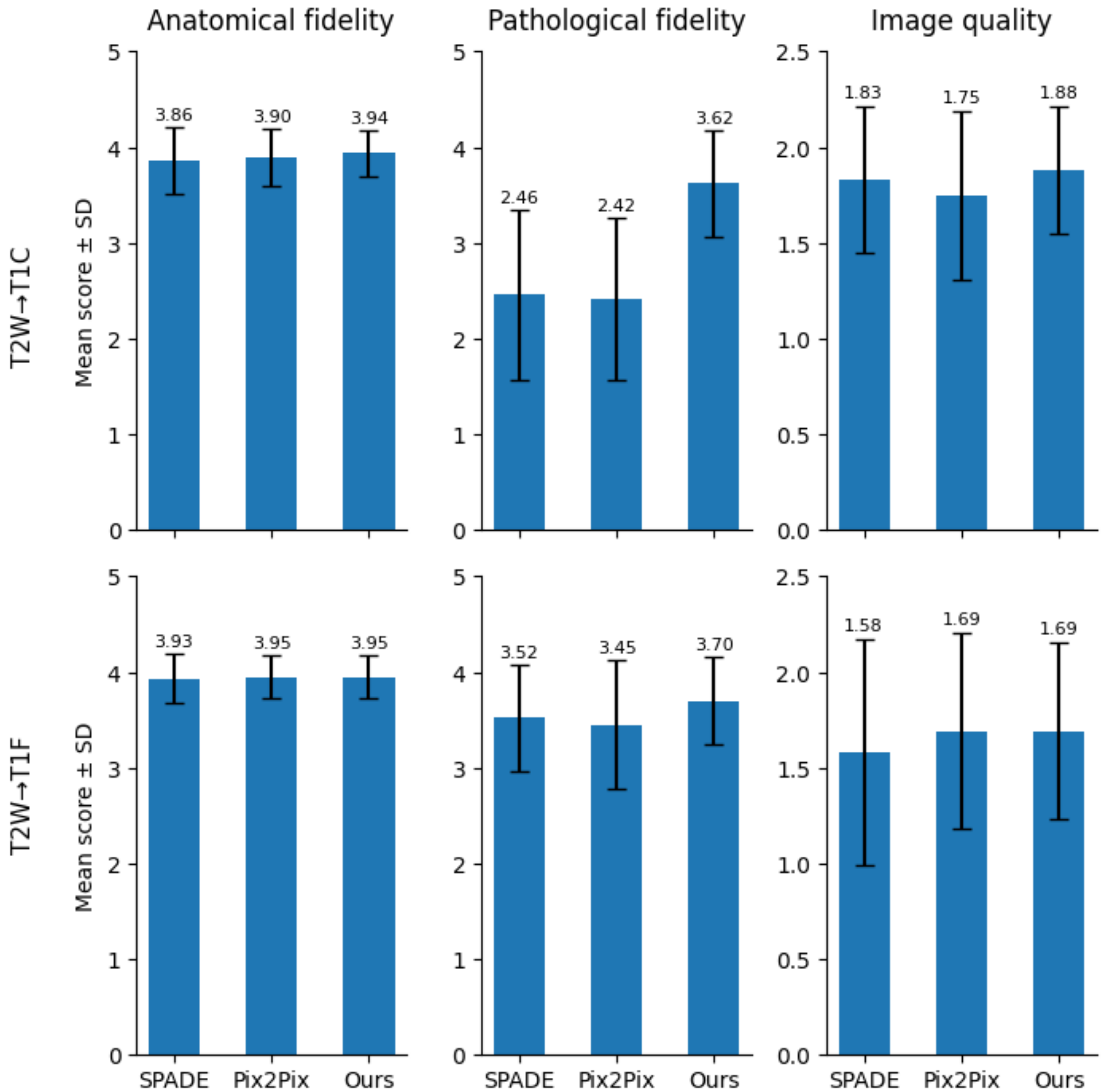

Figure 4: *Radiologists' assessment of the anatomical fidelity, pathological representation, and overall image quality of synthesized MRI sequences.* Scores for *GliomaINK*, Pix2Pix, and SPADE are compared for the evaluated T2W-to-T1C and T2W-to-T2F translation tasks. Results summarize the assessments of two radiologists, including combined scores for anatomical fidelity, pathological representation, and image quality. Higher scores indicate better perceived fidelity or quality according to the predefined evaluation criteria.

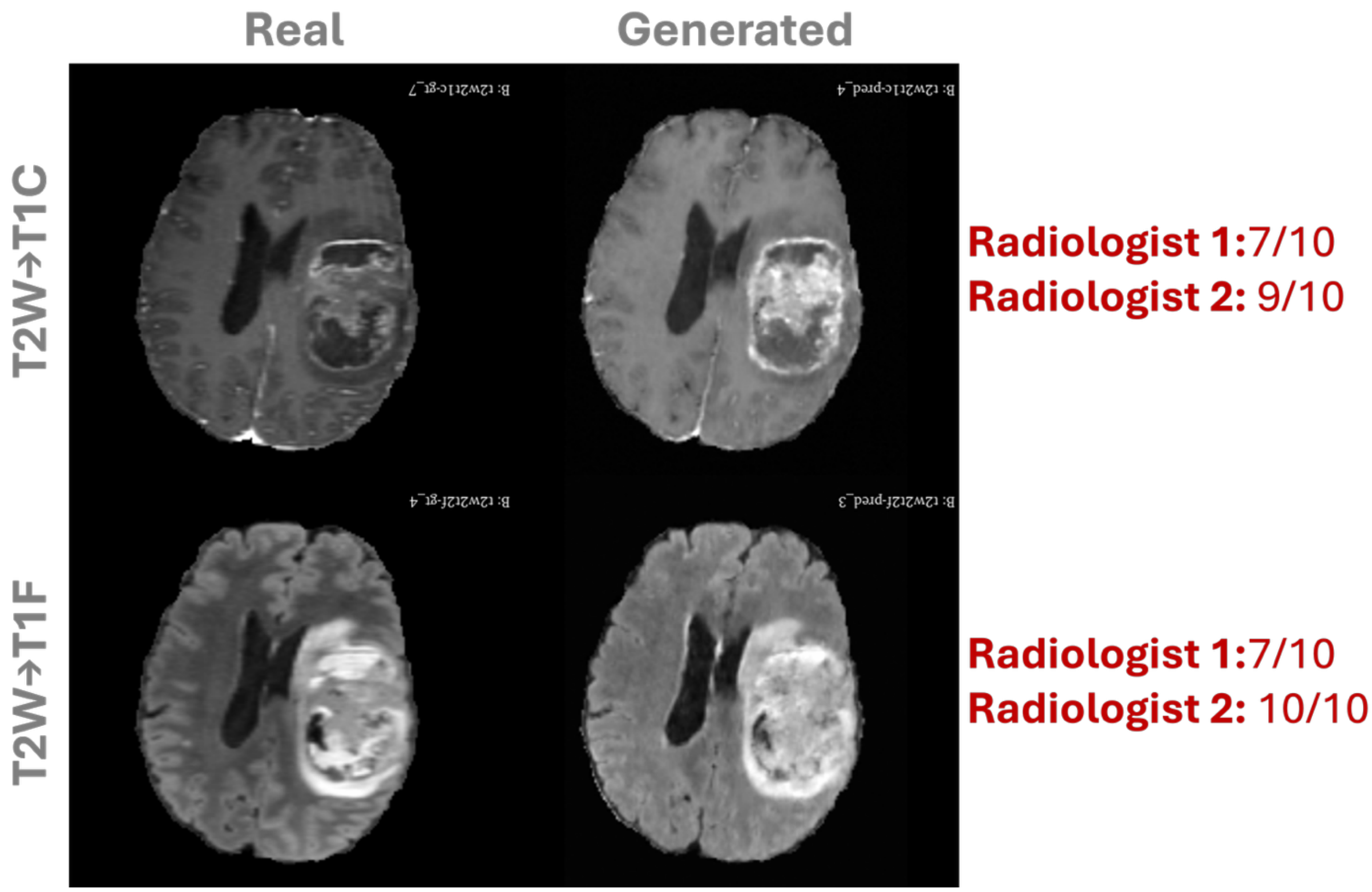

Figure 5: *Inter-observer variability in the qualitative evaluation of synthesized MRI sequences.* Individual scores assigned by the two radiologists are compared for anatomical fidelity, pathological representation, and overall image quality. The comparison illustrates the degree of agreement and variability between readers when evaluating the anatomical and pathological realism of synthesized images.

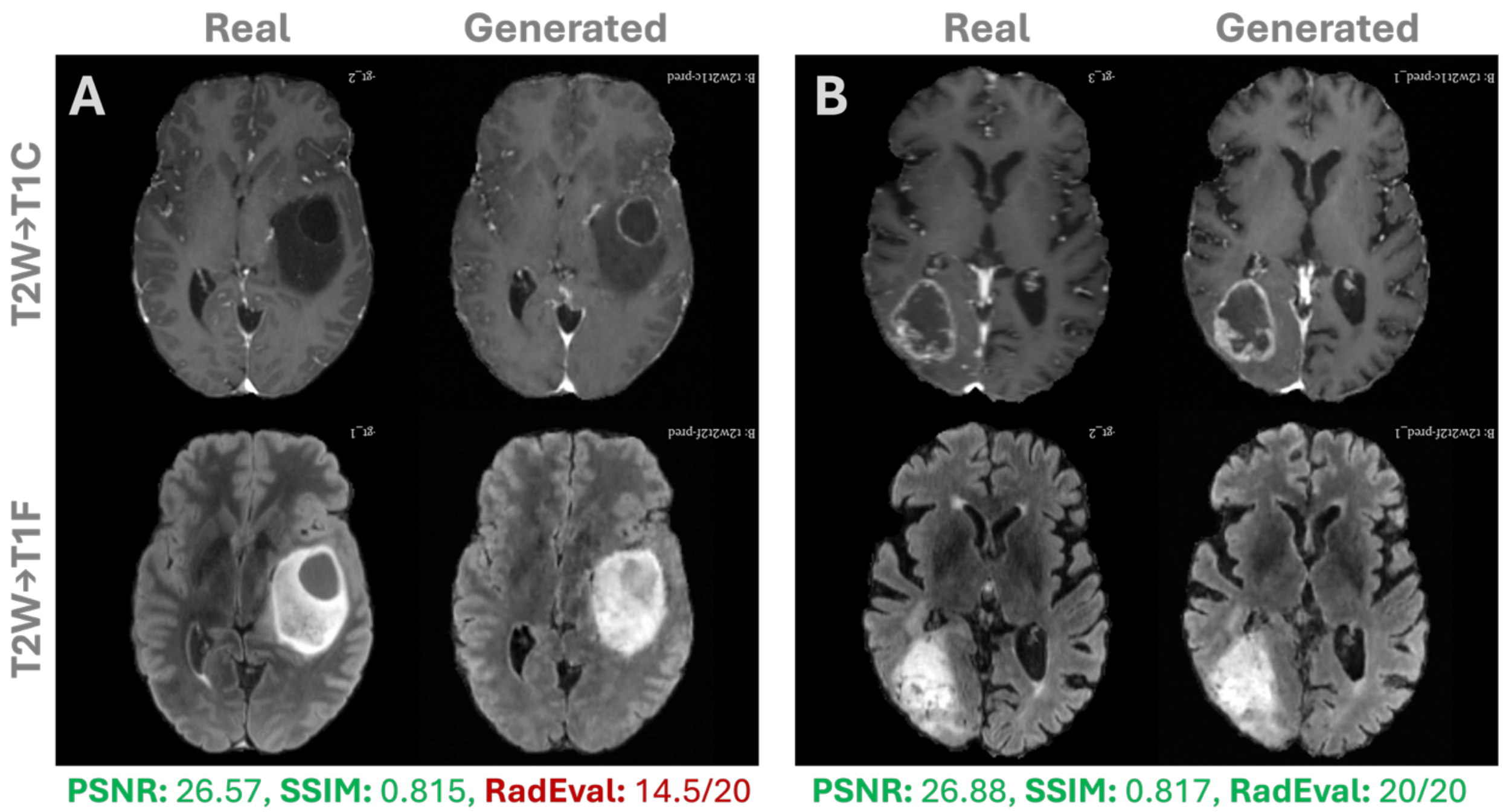

Figure 6: *Comparison of conventional image quality metrics and radiologist evaluations in cross-sequence MRI synthesis.* Case A illustrates a discrepancy between conventional image quality metrics and radiologist assessment: despite favorable PSNR (26.57) and SSIM (0.815), the synthesized images receive a poor radiologist evaluation score (RadEval: 14.5/20), indicating that high pixel-level similarity does not necessarily ensure clinically satisfactory preservation of pathological features and anatomical structures. Case B demonstrates concordant results, with favorable PSNR (26.88) and SSIM (0.817) alongside a high radiologist evaluation score (RadEval: 20/20), indicating good image similarity and radiologist-assessed fidelity. These examples highlight the importance of complementary radiologist evaluation alongside conventional image quality metrics for assessing the clinical reliability of synthetic MRI.

# Appendix

Supplementary Table 1: Radiologist Scoring Criteria for Evaluating Synthesized MRI Images

| Metric | Score | Evaluation criteria |
|---|---|---|
| Anatomy | 0 | Major missing structures or unrealistic anatomy |
| | 1–2 | Some structures misrepresented; noticeable abnormalities |
| | 3 | Mostly realistic with minor flaws |
| | 4 | Fully realistic anatomy |
| Pathology | 0 | No pathology when expected or pathology is unrealistic |
| | 1–2 | Pathology is present but implausible or misplaced |
| | 3 | Convincing pathology with minor limitations |
| | 4 | Pathology looks clinically real and well-integrated |
| Image Quality | 0 | Non-diagnostic (blurry, too noisy, over/underexposed) |
| | 1 | Imperfect but readable |
| | 2 | High-quality, clinically acceptable image |

Supplementary Table 2: Radiologist Assessment of T2W-to-T1C MRI Synthesis.

| **Model** | **Anatomy** | | | **Pathology** | | | **Image Quality** | | |
|---|---|---|---|---|---|---|---|---|---|
| | **Rad. 1** | **Rad. 2** | **Combined** | **Rad. 1** | **Rad. 2** | **Combined** | **Rad. 1** | **Rad. 2** | **Combined** |
| SPADE | 3.78 ± 0.42 | 3.94 ± 0.24 | 3.86 ± 0.35 | 2.12 ± 1.02 | 2.80 ± 0.57 | 2.46 ± 0.89 | 1.86 ± 0.35 | 1.80 ± 0.40 | 1.83 ± 0.38 |
| Pix2Pix | 3.84 ± 0.37 | 3.96 ± 0.20 | 3.90 ± 0.30 | 2.02 ± 0.94 | 2.82 ± 0.52 | 2.42 ± 0.85 | 1.80 ± 0.40 | 1.70 ± 0.46 | 1.75 ± 0.44 |
| *GliomaINK* | **3.90 ± 0.30** | **3.98 ± 0.14** | **3.94 ± 0.24** | **3.48 ± 0.65** | **3.76 ± 0.43** | **3.62 ± 0.56** | **1.88 ± 0.33** | **1.88 ± 0.33** | **1.88 ± 0.33** |

Supplementary Table 3: Radiologist Assessment of T2W-to-T2F MRI Synthesis.

| **Model** | **Anatomy** | | | **Pathology** | | | **Image Quality** | | |
|---|---|---|---|---|---|---|---|---|---|
| | **Rad. 1** | **Rad. 2** | **Combined** | **Rad. 1** | **Rad. 2** | **Combined** | **Rad. 1** | **Rad. 2** | **Combined** |
| SPADE | 3.86 ± 0.35 | **4.00 ± 0.00** | 3.93 ± 0.26 | 3.56 ± 0.50 | 3.48 ± 0.61 | 3.52 ± 0.56 | 1.94 ± 0.24 | 1.22 ± 0.62 | 1.58 ± 0.59 |
| Pix2Pix | **3.92 ± 0.27** | 3.98 ± 0.14 | **3.95 ± 0.22** | 3.54 ± 0.50 | 3.36 ± 0.80 | 3.45 ± 0.67 | **1.96 ± 0.20** | 1.42 ± 0.57 | **1.69 ± 0.51** |
| *GliomaINK* | **3.92 ± 0.27** | 3.98 ± 0.14 | **3.95 ± 0.22** | **3.74 ± 0.44** | **3.66 ± 0.48** | **3.70 ± 0.46** | 1.74 ± 0.44 | **1.64 ± 0.48** | 1.69 ± 0.46 |